\documentclass[nopubldata]{lpor2014}
\graphicspath{{figs/}}
\usepackage{units}
\usepackage{upgreek}
\usepackage{amsmath}
\usepackage{amssymb}
\usepackage{hyperref}
\hypersetup{%
  colorlinks,
  plainpages=false,
  pdfpagelabels,
  breaklinks,
  pdfview=Fit,
  bookmarksopen,
  bookmarksnumbered,
  linkcolor=black,
  anchorcolor=black,
  citecolor=black,
  filecolor=black,
  urlcolor=LPRred
  }%
\usepackage{epstopdf}
\usepackage{color}

\setpages[]{}
\setvolume[0]{0}
\setyear{2012}%
\setdoi{201100000}%
\begin{document}

\titlefigure{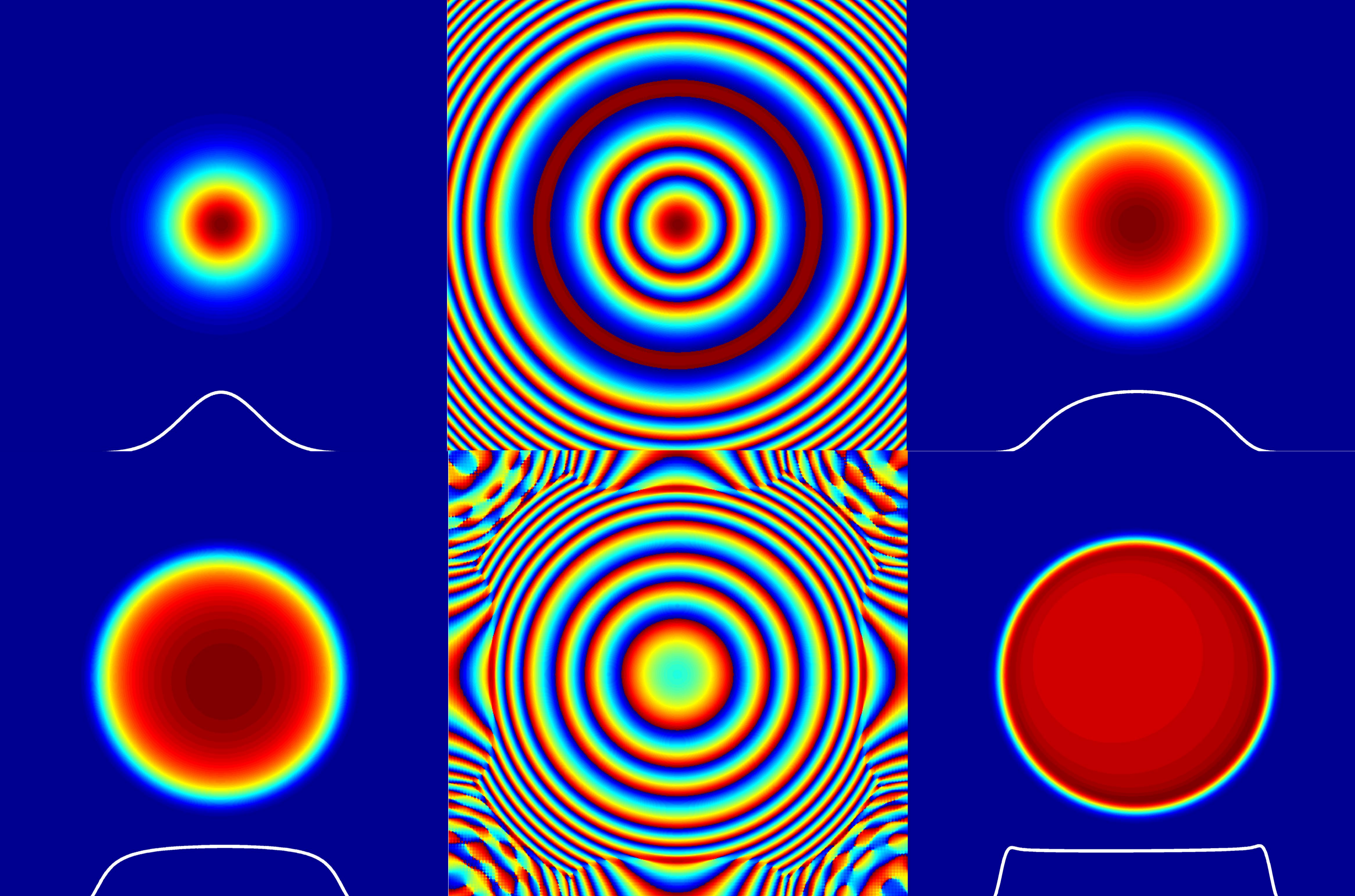}

\abstract{Laser brightness is a measure of the ability to deliver intense light to a target, and encapsulates both the energy content and the beam quality.  High brightness lasers requires that both parameters be maximised, yet standard laser cavities do not allow this.  For example, in solid-state lasers multimode beams have a high energy content but low beam quality, while Gaussian modes have a small mode volume and hence low energy extraction, but in a good quality mode.  Here we overcome this fundamental limitation and demonstrate an optimal approach to realising high brightness lasers. We employ intra-cavity beam shaping to produce a Gaussian mode that carries all the energy of the multimode beam, thus energy extraction and beam quality are simultaneously maximised.  This work will have a significant influence on the design of future high brightness laser cavities.}

\title{Optimised brightness from solid-state lasers}
\titlerunning{High brightness laser}

\author{Darryl Naidoo\inst{1,2}, Igor A. Litvin\inst{1} and Andrew Forbes\inst{1,2}}
\authorrunning{D. Naidoo et. al.}

\institute{%
	CSIR National Laser Centre, P.O. Box 395, Pretoria 0001, South Africa
\and
	School of Physics, University of the Witwatersrand, Private Bag 3, Wits 2050, South Africa
}
\mail{\textsuperscript{*}\,Corresponding author: Darryl Naidoo, e-mail: dnaidoo3@csir.co.za}

\keywords{Laser modes, flat-top beams, laser brightness} 
\maketitle
\section{Introduction}
The brightness of a laser source is a characteristic that encapsulates the energy or power content and the quality of the laser mode. Bright sources are of particular importance in applications where high energy is to be delivered to some distant target, in laser materials processing where high power is required at some specific plane and in long distance free space optical communication. The brightness, $B$, describes the potential of a laser beam to achieve high intensities while maintaining a large Rayleigh range for small focusing angles, which is strongly dependent on the quality of the transverse mode at the output, and is defined as the power $(P)$ emitted per unit surface area $(A)$ per unit solid angle $(\Omega)$. This can be expressed in terms of the beam quality factor, $\textrm{M}^{2}$, as
\begin{equation}
B =\frac{P}{A \Omega} = \frac{P}{\textrm{M}^{4} \lambda^{2}}
\label{eq:1}
\end{equation}

\noindent where ${\textrm{M}^{2}=4\pi \sigma_{x} \sigma_{s_{x}}/\lambda}$, and $\sigma_{x}$ represents the second moment real beam variance corresponding to the time-averaged intensity profile, while $\sigma_{s_{x}}$ corresponds to the spatial frequency distribution and $\lambda$ represents the wavelength of the laser beam. For a laser beam of width, $w$, and angular divergence, $\theta$, that is characterised by circular symmetry, the area of the laser beam may be expressed as $A=\pi w^{2} = 4\pi \sigma^{2}$ and the solid angle expressed as $\Omega=\pi \theta^{2}=4\pi \sigma_{s}^{2}$.  Brightness is therefore proportional to the beam's power (energy) and inversely proportional to the quality of the mode.  

There have been many approaches to improving brightness from laser sources, from the use of fibre lasers \cite{1,2}, to coherent beam combining \cite{3,4,5}, to semiconductor lasers \cite{6,7,8}.  An obvious approach to attempt in realising high brightness in solid-state lasers is through increasing the output laser power by scaling up the input pump power. Doing this is known to increase the thermal load in the laser material giving rise to a non-uniform temperature distribution. In the case of solid-state lasers this results in the generation of phase aberrations which perturb the oscillating mode so that the beam quality degrades. In the worst cases, the increase in power is negated by the decrease in beam quality so that the nett result is a brightness that does not change or worse, actually decreases.  For this reason it is understood that increasing the input pump power is not in general a route to high brightness lasers \cite{9,10,11,12,13}.  Alternatively one may maximise the beam quality factor through exciting a low order mode, but this is usually at the expense of energy extraction due to a smaller gain volume.  Thus conventional solid-state laser paradigms dictate that it is not possible to simultaneously maximise both the mode quality and mode energy at the same time.  

In this Letter we show how a laser cavity can be designed to maximise the mode extraction volume and the mode quality simultaneously, thus optimising the brightness.  Using a diode-pumped solid-state laser as an example, we employ intra-cavity beam shaping optics to force what we call ``mode metamorphosis'': our laser is designed to have a flat-top beam at the gain end of the cavity and a Gaussian beam at the output coupler end.  This approach breaks the standard design paradigm that a mode of a cavity is a particular beam (e.g., Gaussian or flat-top) everywhere.  Here, the mode at any given position repeats after every round trip but changes everywhere along the length of the cavity, from one intensity profile to another.  The consequence of ensuring a flat-top beam at the gain end is that the entire gain volume is used, extracting energy as a if a multimode beam was oscillating, while delivering this energy in a low divergence Gaussian mode at the other end of the cavity:  a Gaussian beam with the energy of a multimode beam.  In this sense this is an optimal approach to realising high brightness solid-state lasers.

\begin{figure}
\centering
\includegraphics[width=8cm]{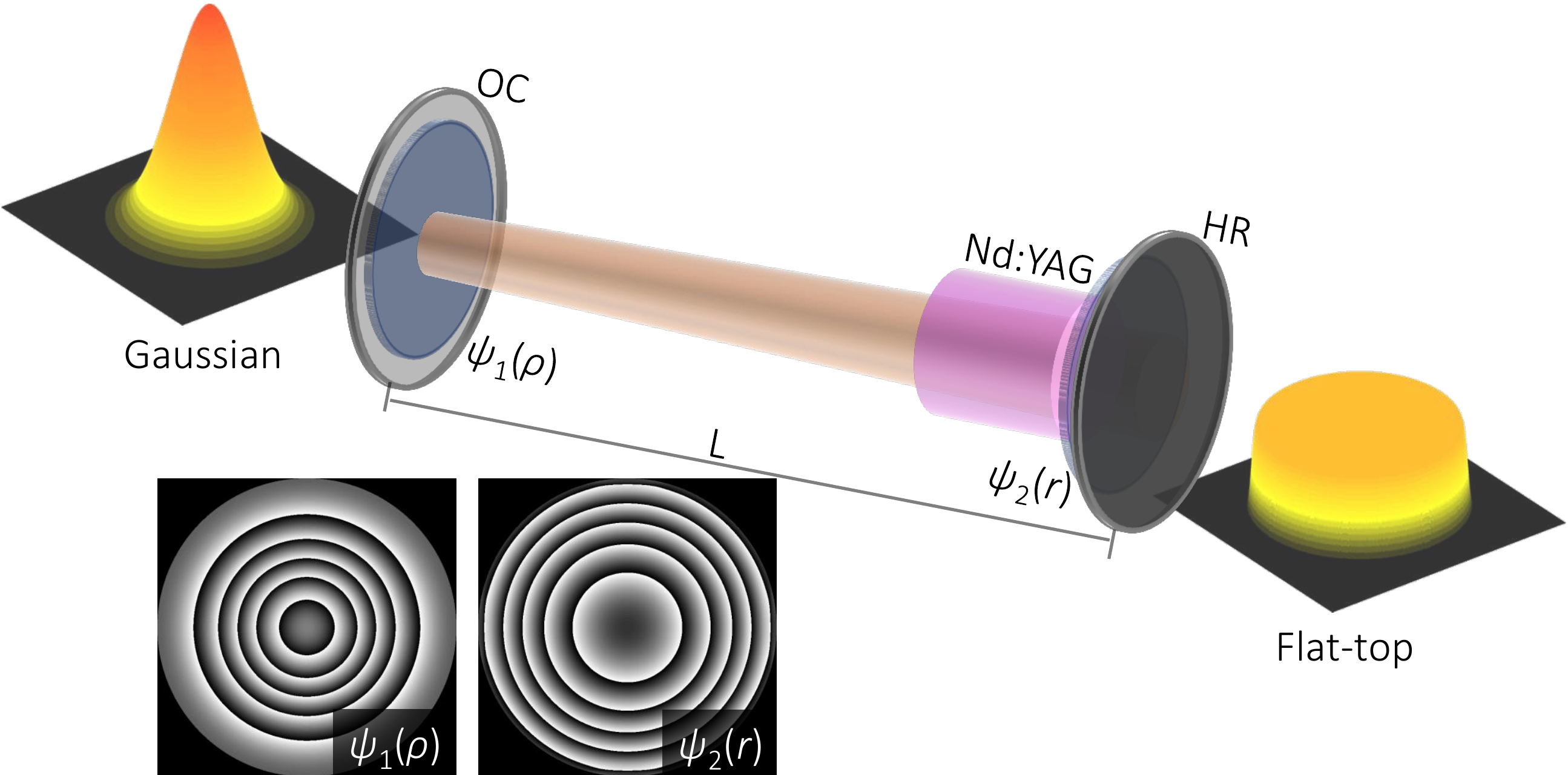}
\caption{Schematic showing the metamorphosis of a Gaussian beam into a flat-top beam with two phase-only optical elements. The first optical element, $\psi_{1}(\rho)$, is used to transform a Gaussian field into a flat-top. The second element, $\psi_{2}(r)$, is encoded as the conjugate of the field at that plane such that the output flat-top beam has a flat wavefront. Both elements have a phase variation of 0 to 2$\pi$ (white to black) and are depicted as grayscale images.}
\label{fig:1}
\end{figure}

\section{Concept}

Our idea is based on the ability to create a mode that is not the same everywhere in the cavity yet repeats at each position after one round trip.  In particular, we wish to create a cavity with a Gaussian mode at one end and a flat-top beam at the other.  This is a well-known beam shaping problem outside the laser cavity \cite{14,15,16,17,18,19,20,21}.  By employing a cavity with two beam shaping elements, we are able to have a mode that changes continuously in shape during propagation from one mirror to the other, morphing from a Gaussian profile to a flat-top profile.  An important aspect is that we use two phase-only elements for the transformation; this allows the beam shaping to be done in a lossless manner, thus minimising cavity losses.  It is also possible to design the cavity with one beam shaping element but at the expense of employing complex amplitude modulation, i.e., loss of amplitude and thus higher cavity loss.  

To design our two phase-only optics we consider the desired beam at the output coupler (OC) as in Fig.~\ref{fig:1} to be a Gaussian field with a flat wavefront of the form $u_{G}\left(\rho\right)=\exp[-\left(\rho/w_{0}\right)^{2}]$, where $w_{0}$ is the Gaussian beam width (at the output coupler).  We wish to transform this into a flat-top beam (FTB) at the gain end with a profile of $u_{FTB}\left(\rho\right)=\exp[-\left(\rho/w_{FTB})^{2N}\right]$ with $N>1$ defining the steepness of the flat-top edges.  If the element at the OC is comprised of a Fourier transforming lens and some phase-only transmission component, $\psi_{F}$, and if the optical length of the cavity matches that of the Fourier transforming lens ($L=f$), then we may determine an analytical solution for the phase function of the first element through the stationary phase approximation \cite{21}:

\begin{equation}
\psi_{1}\left(\rho\right)=\psi_{F}\left(\rho\right)-\frac{k\rho^{2}}{2f},
\label{eq:6}
\end{equation}
with the term $k\rho^{2}/2f$ is required for the lens and 

\begin{equation}
\psi_{F}\left(\rho\right)=\alpha\frac{\sqrt{\pi}}{2}\int\limits^{\rho/w_{0}}_{0} \sqrt{1-\exp\left(-\xi^{2}\right)}d\xi,
\label{eq:4}
\end{equation}
where $\alpha$ is a dimensionless parameter that is defined as
\begin{equation}
\alpha=\frac{2\pi w_{0} w_{FTB}}{f \lambda}.
\label{eq:5}
\end{equation}

As with the phase profile of the first element, $\psi_{1}$, it is also possible to apply the stationary phase approximation to determine a closed form solution for the phase profile of the second element at the plano high reflecting (HR) mirror as
\begin{equation}
\psi_{2}\left( r \right)=\textrm{arg}\left\lbrace\exp\left[i\left(\frac{kr^{2}}{2f}+\psi_{F}\left(\vartheta\left(r\right)\right)-\frac{\alpha r \vartheta\left(r\right)}{w_{0} w_{FTB}}\right)\right]\right\rbrace,
\label{eq:7}
\end{equation}
where the unknown function $\vartheta(r)$ may be determined from the stationary phase condition $r/w_{FTB}=\partial\psi_{F}/\partial\rho$. This ensures that the flat-top output has a flat wavefront that when reverse propagated, due to the reciprocity of light, morphs into a Gaussian beam at the OC.

\begin{figure}
\centering
\includegraphics[width=6cm]{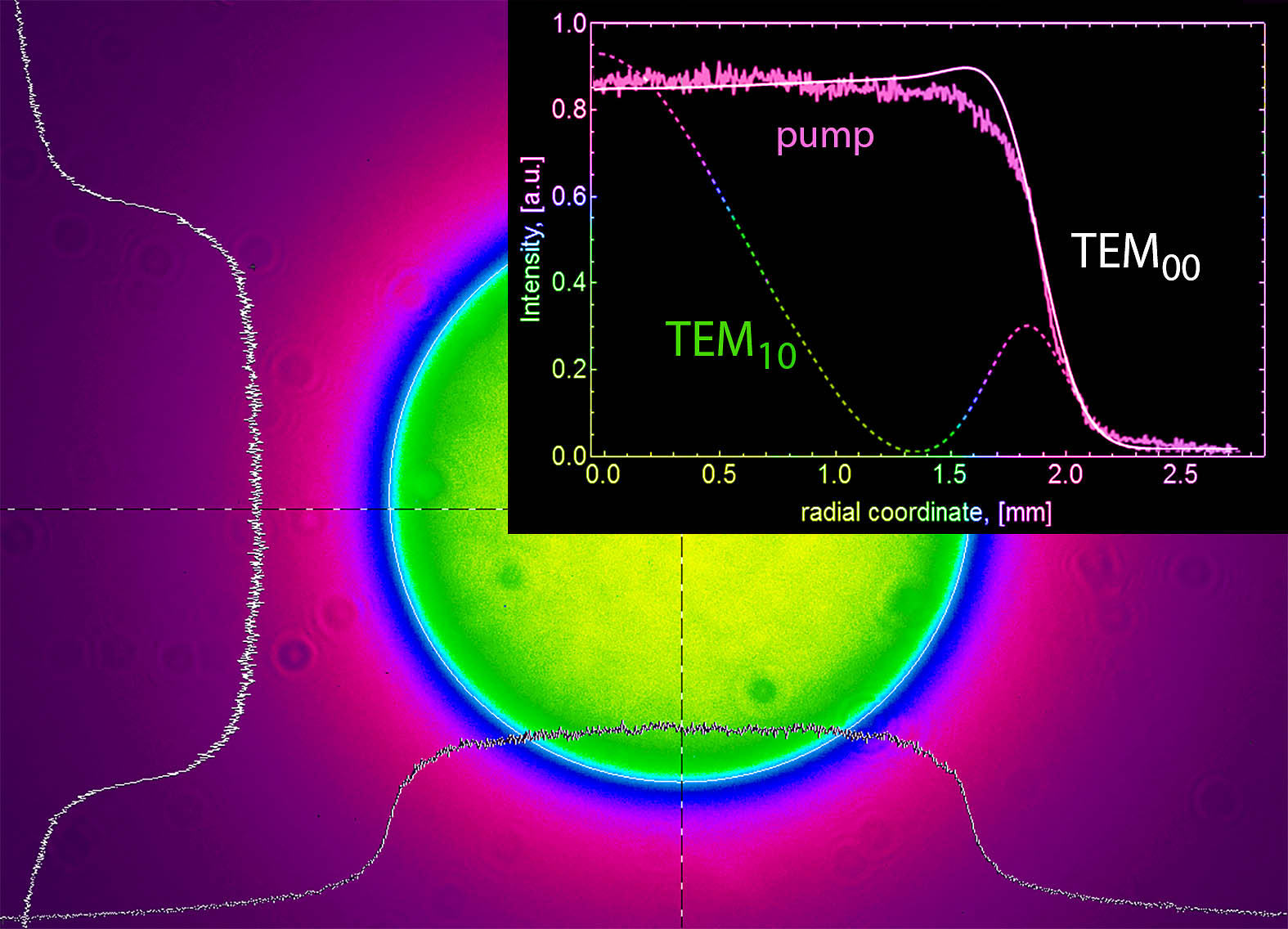}
\caption{The modal discrimination of higher unwanted modes is determined primarily by the pump to mode overlap and not by the modal diffraction losses.  Here we show the fluorescence of the side pumped gain medium used in our experiment, which closely approximates as flat-top.  The overlap between this gain and the first two lowest loss modes in the cavity, TEM$_{00}$ and TEM$_{10}$ respectively, is given by $\eta_{00}$ = 99.6$\%$ and $\eta_{10}$ = 77.2$\%$.}
\label{fig:2}
\end{figure}

\begin{figure*}[htbp]
\centering
\includegraphics[width=16cm]{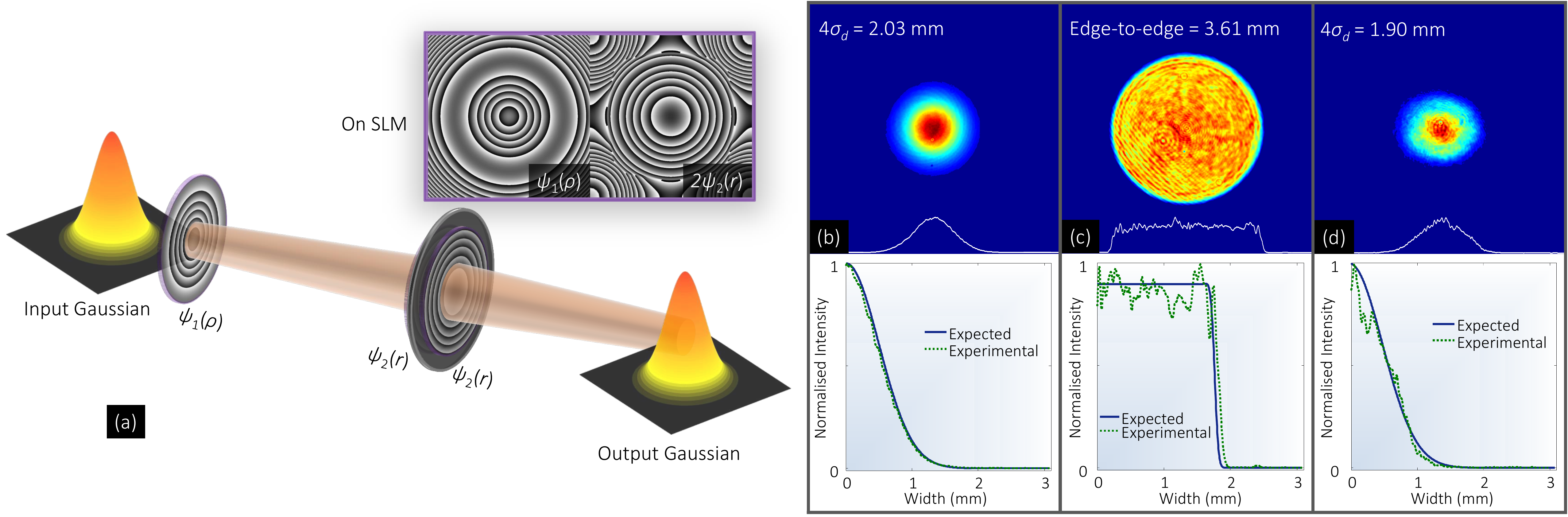}
\caption{(a) A single intra-cavity round trip may be represented as an unfolded cavity with two phase transformations. The external optical testing of a single round trip was executed by propagating a collimated (b) Gaussian beam of $w_{0}=1~\textrm{mm}$ onto an appropriate phase pattern for transformation to a flat-top beam of width $w_{FTB_2}=1.8~\textrm{mm}$. The output resulted in well defined (c) flat-top beam of edge-to-edge diameter of 3.61 mm (measured $w_{FTB_2} = 1.81~\textrm{mm}$) . The flat-top beam was directed onto the second element and was transformed to a (d) Gaussian beam of measured $w_{0} = 0.95~\textrm{mm}$. The profiles below the 2D beam images demonstrate high overlap between the expected intensities and the experimentally measured intensities.}
\label{fig:3}
\end{figure*}

Note that the profiles at the gain end and output coupler end can be designed arbitrarily.  In the case of our example of a side pumped solid-state laser, we select the profile at the gain end to match the gain profile. The use of this approach allows for efficient mode discrimination of unwanted higher order modes. In such a cavity the discrimination is not seen in the diffraction losses but rather in the pump to mode overlap, as the difference in the diffraction losses for the first six competing modes is less than one percent (as calculated by a Fox-Li analysis of the cavity).  If we take into consideration the overlap of the gain and the mode by considering the fluorescence of the gain, then we find a significant difference in eigenvalues of the first two competing modes, namely TEM$_{00}$ = 99.6$\%$ and TEM$_{10}$ = 77.2$\%$ as illustrated in Fig.~\ref{fig:2}). This demonstrates that the diffraction losses are insignificant in the selection of the output mode, however, the pump to mode overlap of the fundamental mode provides the required modal discrimination.

\section{Results}

The concept of the laser resonator as presented in Fig.~\ref{fig:1} was optically tested external to the laser cavity in an unfolded geometry using spatial light modulators (SLM) to encode the desired phases. The unfolded round trip simulates the propagation of the cavity mode from the OC to the HR and back along the same path. A single round trip requires four optical phase transformations as we pass through each optical element twice, as illustrated in Fig.~\ref{fig:3}~(a). The intensity profiles immediately before and after each element are identical, although their phase compositions are markedly different. This implies that on the return pass (HR to OC) the intensity of the field immediately before $\psi_{1}(\rho)$ may be adequately analysed to infer the intensity after the optical element.

We addressed a spatial light modulator (SLM) with phase profiles, $\psi_{1}(\rho)$ and $2 \psi_{2}(r)$, one to each half of the SLM, using a split screen functionality. We directed a collimated Gaussian beam operating at $\lambda=633~\textrm{nm}$ with $w_{0} = 1~\textrm{mm}$ onto the right half of the SLM. This half was addressed with a grayscale phase pattern of the first element, which contained an encoded lens of $f=250~\textrm{mm}$ for the selection of a flat-top beam of $2w_{FTB} = 3.6~\textrm{mm}$. The output resulted in a well defined flat-top beam at 250 mm from the plane of the SLM with a measured edge-to-edge diameter of 3.61 mm as shown in Fig.~\ref{fig:3}~(c).

The flat-top beam was directed by a flat mirror and propagated the cavity length to the left half of the SLM which was addressed with a grayscale phase pattern of the second element with double its phase to simulate a double pass. The output field resulted in a Gaussian beam (see Fig.~\ref{fig:3}~(d)) at 250 mm from the SLM plane with a diameter of 4$\sigma$ = 1.90 mm (measured $w_{0} = 0.95~\textrm{mm}$). This external test with SLMs served to confirm that the design principle works, and that while the cavity mode is always changing, it is nevertheless repeated after each round trip.  The measured and calculated profiles were all in excellent agreement, both in terms of profile shape and size, as seen in Fig.~\ref{fig:3}.  We do notice some noise overlaid with our desired profiles which could be due to pixelation of the SLMs or the phase wrapping of the phase functions. Such noise has high spatial frequency and would be expected to contribute to small additional losses inside the cavity. 

\begin{figure*}
\centering
\includegraphics[width=16cm]{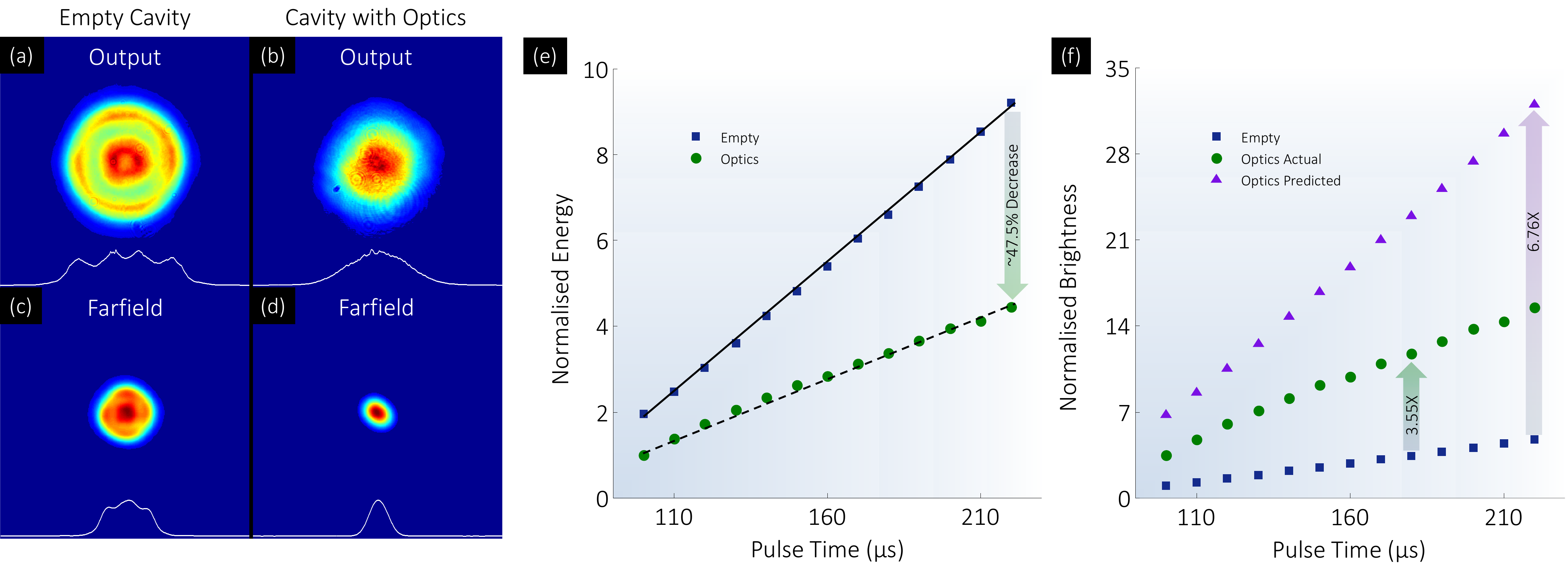}
\caption{The laser cavity with the inclusion of the beam transformation elements was compared to an empty cavity. Near field outputs from the two cavities show (a) multimode operation for the empty cavity and (b) single mode Gaussian-like operation for the custom cavity.  The far-field profiles likewise confirm this property, with the empty cavity shown in (c) and the custom cavity in (d).  (e) The output energy of the cavity with the optical transformation elements was measured and normalised to unity for comparison to the empty cavity. (f) The brightness for the cavity with the optics as compared to the empty cavity increased by $3.55\times$ while with astute control of the flat-top beam diameter, the cavity may be engineered to have a brightness improvement of $6.76\times$.  The input energy was adjusted by varying the pulse time of the input excitation.}
\label{fig:4}
\end{figure*}

The experimental implementation of the intra-cavity metamorphosis of a Gaussian beam to a flat-top beam was realised in a diode-pumped solid-state laser where a 0.5 at.-$\%$ Nd-doped YAG rod (4 mm $\times$ 50 mm) was side pumped with a total input average pump power of $\sim 600$ W where the input pump energy was pulsed at 20 pulses per second (20 pps). The cavity mirrors were both plane with the back mirror acting as a high reflector (HR) while the output coupler (OC) had a reflectivity of $40\%$ as is illustrated in the experimental schematic in Fig.~\ref{fig:1}. The optical elements were positioned sufficiently adjacent to the mirrors and were suitably mounted for accurate control of the lateral positioning and pitch and yaw. The length ($L$) of the cavity was adjusted to accommodate the gain medium and was increased from its nominal length of 538 mm (corresponding to the embedded lens of $f=538~\textrm{mm}$) to 570 mm.  To design diffractive optics based on this cavity, we opted for a Gaussian beam of $w_{0} = 1.5~\textrm{mm}$ with a transformation into a flat-top beam of $w_{FTB} = 1.75~\textrm{mm}$. This was to ensure minimal diffraction effects, with beam widths much smaller than the cavity apertures sizes, which came at the expense of energy extraction (see later discussion). With these parameters, the phase profiles of the two elements under these specifications were computed and were used in the manufacturing of physical phase-only elements.

The key point of comparison to the resonator as presented in Fig.~\ref{fig:1}, was an optical resonator of the same cavity length and output mirrors but without the inclusion of the phase-only optical elements.  We refer to this as the ``empty cavity''.  The output of the empty cavity at 20 pps is illustrated in Fig.~\ref{fig:4}~(a) and had a measured second moment beam diameter of 4$\sigma$ = 3.45 mm. It is clear from the intensity cross-sections that the output is not single mode and this is further verified by sampling the intensity of the output in the Fourier plane of the output coupler as shown in Fig.~\ref{fig:4}~(c). With the insertion of the optical elements in the cavity, the output at 20 pps is approximately Gaussian in shape as illustrated in Fig.~\ref{fig:4}~(b) with a measured second moment beam diameter of 4$\sigma$ = 2.95 mm which is in excellent agreement with the design parameter of $2w_{0} = 3~\textrm{mm}$ . The Gaussian shaped output was further validated in its far field intensity profile as shown in Fig.~\ref{fig:4}~(d).

The output of the empty cavity and the cavity with the optics inserted were analysed in terms of the beam quality (M$^{2}$) and output energy, which were used to determine the brightness at the output. An ISO-compliant method was used to determine the M$^{2}$ which resulted in $\textrm{M}^{2}= 4.03$ for the empty cavity. With the optics inserted in the cavity the measured M$^{2}$ improved dramatically as compared to the empty cavity and resulted in $\textrm{M}^{2}= 1.55$, confirming that the output was a lower divergence Gaussian-like beam.

In determining the brightness at the output, we measured the output energy of the cavities as is illustrated in Fig.~\ref{fig:4}~(e). Because the gain volume of the two cavities were not the same; their extraction was markedly different.  The small flat-top beam meant a designed gain volume of only half that of the empty cavity, with an anticipated $50\%$ loss in energy as the flat-top beam diameter traversing the length of the gain medium was half of that of the gain medium.  The measured extracted energy was $\sim 47.5\%$ as compared to the empty cavity, with the additional $\sim 2.5\%$ loss attributed to losses imposed by the optical elements, length optimisation and alignment.  We point out that the two cavities could be designed for identical energy extraction by increasing the flat-top beam size appropriately (cost of the fabrication of new elements prohibited us from doing this).

The measured output energy and beam quality of the cavities were used to determine the optical brightness from Eq.~(\ref{eq:1}). Without any adjustment for gain volume and comparing the performances directly, we find  an increase in laser brightness by over 350$\%$, as shown in Fig.~\ref{fig:4}~(f). The increase in optical brightness illustrates a marked improvement on the empty cavity and may be further improved by exploring a larger flat-top beam over the length of the crystal. The conservative flat-top beam diameter chosen in this investigation may be increased to match the size of the multi-mode beam ($\sim 3.5$ mm) and in doing so, we predict that the energy extraction may be equivalent to that of the empty cavity. This potentially results in an increased optical brightness of close to 700$\%$ (a factor of $6.76\times$) as illustrated in Fig.~\ref{fig:4}~(f).  

\section{Discussion}
The salient point here is that the cavity under study was an off-the-shelf commercial laser and was not adjusted in any way other than to insert two custom designed phase-only optical elements.  The effect of doing this was to dramatically enhance the performance in terms of output brightness, possible through the concept of mode metamorphosis inside a laser cavity.  We believe that with careful selection of the parameters and better alignment of the optics that an order of magnitude brightness is possible from an otherwise standard cavity.  In particular this concept may be revolutionary for slab laser designs.  The $\textrm{M}^{2}$ in the long axis can be well over 100.  Applying this principle to such a cavity would require a 1D rather than 2D beam shaping solution, and would have a predicted brightness enhancement of several orders of magnitude: high power Gaussian beams from slab lasers.  Finally, while we have outlined the concept with solid-state lasers it is clear that the principle is general enough to apply to a much wider variety of cavities.
 
\section{Conclusion}
In this Letter we have demonstrated a high brightness solid-state laser that optimises both energy extraction and beam quality, two parameters that are usually anti-correlated, i.e., increasing one tends to decrease the other.  We achieve this by introducing a new design paradigm whereby the laser mode inside the cavity changes continuously from one desired shape to another.  By specifying a flat-top beam at the gain end and a Gaussian beam at the output coupler end, the output from the cavity has the energy of a multimode beam but in a low divergence Gaussian beam.  We point out that this concept is not restricted to these two particular intensity profiles and may be suitably adapted for alternative geometries, e.g., rectangular beams for a slab configuration or annular beams for annular gain.  Our demonstration therefore serves as a general approach to optimising laser brightness from laser cavities.



\end{document}